\documentclass{elsart5}

 \usepackage{graphicx}

\usepackage{amssymb}

\begin{document}

\begin{frontmatter}

\title{Magentic-Field Induced Quantum Phase Transition and Critical Behavior in a Gapped Spin System TlCuCl$_3$}

\author[aff1]{F. Yamada\corauthref{cor1}}
\ead{yamada@lee.phys.titech.ac.jp}
\corauth[cor1]{}
\author[aff1]{T. Ono}
\author[aff1]{M. Fujisawa}
\author[aff1]{H. Tanaka}
\author[aff2]{T. Sakakibara}
\address[aff1]{Department of Physics, Tokyo Institute of Technology, Tokyo152-8551}
\address[aff2]{Institute for Solid State Physics, The University of Tokyo, Kashiwa, Chiba 277-8581}
\received{12 June 2005}
\revised{13 June 2005}
\accepted{14 June 2005}


\begin{abstract}
Magnetization measurements were performed on TlCuCl$_3$ with gapped ground state. The critical density and the magnetic phase diagram were obtained. The interacting constant was obtained as $U/k_{\rm B} = 313$ K. The experimental phase boundary for $T < 5$ K agrees perfectly with the magnon BEC theory based on the Hartree-Fock approximation with realistic dispersion relations and $U/k_{\rm B} = 320 $ K. The exponent $\phi$ obtained with all the data points for $T < 5$ K is $\phi = 1.99$, which is somewhat larger than theoretical exponent $\phi_{\rm BEC} =3/2$. However, it was found that the exponent converges at $\phi_{\rm BEC} =3/2$ with decreasing fitting window.
\end{abstract}


\end{frontmatter}

\section{Introduction}\label{}
TlCuCl$_3$ is an $S = 1/2$ interacting dimer system, which has an excitation gap $\Delta = 7.5$ K \cite{Shiramura}. In an external magnetic field, the gap closes and the system undergoes magnetic phase transition for $H > 5.4$ T. This phase transition is a magnetic quantum phase transition and is described as the Bose-Einstein condensation (BEC) of spin triplets called magnons or triplons \cite{Nikuni}. 
The BEC theory based on the Hartree-Fock (HF) approximation with a parabolic isotropic dispersion relation ${\hbar}^2 {\mathbf{k}}^2/{2m}$ gives the phase boundary described by the power law 
$\left( g/2 \right) \left[H_{\rm N}(T)-H_{\rm c}\right] \propto T^{\phi} $
 with exponent $\phi_{\mathrm{BEC}}=3/2$ \cite{Nikuni}, where $H_{\rm c}={\Delta}/g{\mu}_{\rm B}$ is the gap field. A point given by $T=0$ and $H=H_c$ on the temperature vs field diagram denotes the quantum critical point. In the previous study \cite{O_mag,O_heat}, it was shown that the experimental phase boundary is expressed by the power law with exponent $\phi=2.0 \sim 2.2$. This exponent is somewhat larger than $\phi_{\mathrm{BEC}} = 3/2$. 
Recently, the deviation of the exponent $\phi$ toward larger value from $\phi_{\mathrm{BEC}} = 3/2$ was theoretically discussed \cite{Kawashima,Nohadani,mo}. 
Misguich and Oshikawa \cite{mo} extended the HF calculation by Nikuni $et \ al.$ \cite{Nikuni} by using a realistic dispersion relation and achieved remarkable quantitative agreement with the experimental phase diagram. They also calculated the critical density of triplons $n_\mathrm{cr}(T)$ and estimated the interacting constant $U/k_{\rm B} = 340 \mathrm{K}$. We carried out magnetization measurements on TlCuCl$_3$ to reevaluate $n_\mathrm{cr}(T)$, $U$ and critical exponent $\phi$.


\section{Experimental details}
Single crystals of TlCuCl$_3$ were grown by the vertical Bridgman method. The  details of preparation were reported in reference\cite{O_mag}.
The magnetization measurements were performed using SQUID magnetometer (Quantum Design MPMS XL) in the temperature region $1.8 \mathrm{K} \leq T \leq 100 \mathrm{K}$ in magnetic fields of up to 7 T. The magnetic fields were applied parallel to the $b$-axis and [2, 0, 1] direction and perpendicular to the (1, 0, $\overline{2}$) plane.
The magnetization measurements were also performed using Faraday Force Magnetometer \cite{Faraday} at Institute of Solid State Physics in the temperature region $30 \mathrm{mK} \leq T \leq 4$ K in magnetic fields up to 8 T. The magnetic fields were applied perpendicular to the (1, 0, $\bar{2}$) plane.

 \begin{figure}[t]
\includegraphics[scale =0.63]{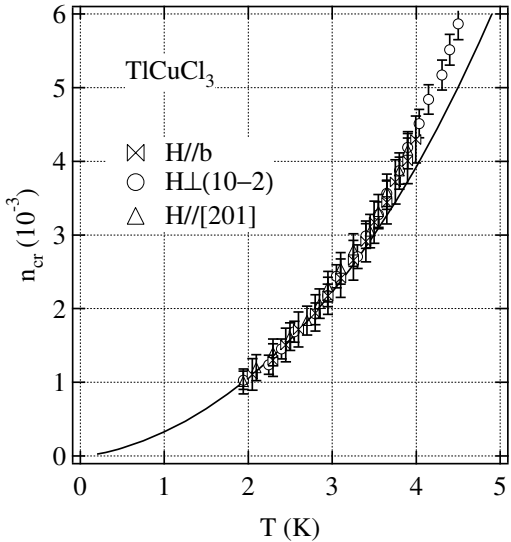}
\caption{Temperature dependence of the critical density of triplons in TlCuCl$_3$ for $H \| b$, $H \| [2, 0, 1]$ and $H \bot  (1, 0,  \overline{2})$. The solid line is the theoretical calculation by Misgich and Oshikawa \cite{mo}.}
 \label{nc}
 \end{figure}

\begin{figure}[t]
\includegraphics[scale =0.63]{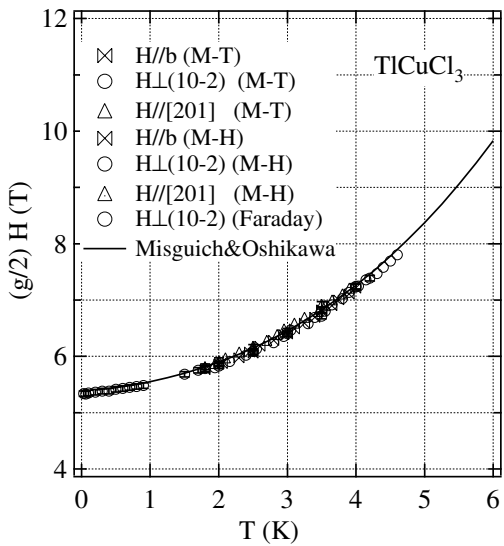}
\caption{Phase diagram normalized by the $g$-factors in TlCuCl$_3$. The solid line is the result of calculation by Misuguich and Oshikawa \cite{mo} }
 \label{phase}
 \end{figure}


\section{Results and discussion}
The critical density $n_{\rm cr}$ of triprons corresponds to the absolute values of the magnetization at $T_\mathrm{N}$ or $H_\mathrm{N}$. Figure \ref{nc} shows the critical density $n_{\rm cr}$ as a function of temperature obtained from the magnetization measurement. 
The solid line in Fig. \ref{nc} is the critical density calculated with the realistic dispersion by Misguich and Oshikawa with $U/k_{\rm B} = 320 $ K \cite{mo}. The experimental and theoretical critical density agree well in the low temperature region for $T \leq 3 \mathrm{K}$. 
From the relation $(g/2) [H_{\mathrm N} (T) - H_{\mathrm c}] = 2Un_\mathrm{cr} \mathrm{(T)} $, the interacting constant $U$ is estimated as $U/k_\mathrm{B} = 312$ K, 311 K and 315 K for $H \| b$, $H \bot  (1, 0,  \overline{2})$ and $H \| [2, 0, 1]$, respectively. Their average is $U/k_\mathrm{B} = 313$ K.

Transition temperatures $T_{\rm N}(T)$ and transition field $H_{\rm N}(T)$ measured by temperature ($M-T$) and field ($M-H$) scans are summarized in Fig. \ref{phase}. Phase boundaries for $H \| b$, $H \| [2, 0, 1]$ and $H \bot  (1, 0,  \overline{2})$ coincide when normalized by the $g$-factor and can be expressed by the power law. The solid line in Fig. \ref{phase} is the result of the BEC theory based on the HF approximation with the realistic dispersion relations and $U/k_{\rm B} = 320 $ K \cite{mo}. Both results agree perfectly for $T < 5$ K. 

We analyze the phase boundary for the magnetic field perpendicular to the $(1,0,\bar 2)$ plane using the power law. In the present analysis, we set the lowest temperature at $T_{\rm min}=30$ mK and varying temperature range for fitting.
Using all the data points , we obtained $\phi = 1.99$, which is somewhat larger than $\phi_\mathrm{BEC} = 3/2$ derived by the triplon BEC theory based on HF approximation \cite{Nikuni} as obtained in the previous measurements $\phi = 2.0 \sim 2.2 $ \cite{O_mag,O_heat}. However, the critical exponent $\phi$ decreases with decreasing fitting window, and converges at $\phi_\mathrm{BEC} = 3/2$, as predicted by Nohadani {\it et al.} \cite{Nohadani}. For $T \leq 1.80$ K, we obtain $\phi = 1.52 \pm 0.06$, which is nearly equal to $\phi_\mathrm{BEC} = 3/2$ derived from the triplon BEC theory \cite{Nikuni}. 

\section{Conclusion}
We have presented the critical density of triplons and the magnetic phase diagram of TlCuCl$_3$. The interacting constant was estimated as $U/k_B = 313 $K.The phase boundary is expressed by the power law and agrees perfectly with the triplon BEC theory based on the HF approximation with realistic dispersion relations and $U/k_{\rm B} = 320 $ K \cite{mo}. The critical exponent $\phi$ decreases with decreasing fitting range. We obtained $\phi = 1.52 \pm 0.06$ for $T \leq 1.80$ K, which is nearly equal to $\phi_\mathrm{BEC} = 3/2$ derived from the triplon BEC theory \cite{Nikuni}. These results strongly support the BEC description of field-induced magnetic ordering in TlCuCl$_3$.

\section{Acknowledgements}
This work was supported by The 21st Century COE Program at Tokyo Tech. "Nanometer-Scale Quantum Physics" both from the Ministry of Education, Culture, Sports, Science and Technology of Japan.


\begin{thebibliography}{00}
\bibitem{Shiramura} W. Shiramura, K. Takatsu, H. Tanaka, M. Takahashi, K. Kamishima, H. Mitamura and T. Goto  {\it J. Phys. Soc. Jpn.} 66 (1997) p.1900 
\bibitem{Nikuni}T. Nikuni, M. Oshikawa, A. Oosawa, and H. Tanaka {\it Phys. Rev. Lett.} 84 (2000), p. 5868.
\bibitem{O_mag} A. Oosawa, M. Ishii and H. Tanaka  {\it J.Phys. : Condens. Matter} 11 (1999) p.265
\bibitem{O_heat} A. Oosawa, H. Aruga Katori and H. Tanaka {\it Phys. Rev. B} 63 (2001) 134416 
\bibitem{Kawashima}N. Kawashima {\it J. Phys. Soc. Jpn.} 79 (2004), p. 3219.
\bibitem{Nohadani}O. Nohadani, S. Wessel, B. Normand, and S. Hass {\it Phys. Rev. B} 69 (2004), 220402(R).
\bibitem{mo} G. Misguich and M. Oshikawa {\it J. Phys. Soc. Jpn.} 73 (2004), p. 3429.
\bibitem{Faraday} T. Sakakibara, H. Mitamura, T. Tayama and H. Amitsuka {\it Jpn. J. Appl. Phys.} 33 (1994) p.5067




\end{thebibliography}
\end{document}